\documentclass{article}
\usepackage{arxiv}

\usepackage[utf8]{inputenc} 
\usepackage[T1]{fontenc}    
\usepackage{hyperref}       
\usepackage{url}            
\usepackage{booktabs}       
\usepackage{amsfonts}       
\usepackage{nicefrac}       
\usepackage{microtype}      
\usepackage{lipsum}
\usepackage{graphicx}
\graphicspath{ {./} }
\usepackage{amsmath}
\usepackage[scr=rsfs]{mathalpha}
\usepackage{subcaption}

\usepackage[authoryear,round]{natbib}
\title{Numerically Consistent Non-Boussinesq Subgrid-scale Stress Model with Enhanced Convergence}

\author{
}

\usepackage{authblk}

\author[1,*]{Yuenong Ling}
\author[1,2]{Adrian Lozano-Duran}

\affil[1]{Department of Aeronautics and Astronautics, Massachusetts Institute of Technology, Cambridge, USA}
\affil[2]{Graduate Aerospace Laboratories, California Institute of Technology, Pasadena, USA}
\affil[*]{Author to whom any correspondence should be addressed.}

\begin{document}
\maketitle

\begin{abstract}
We extend the data-assimilation approach of \cite{ling2025numerically}
to develop machine-learning-based subgrid-scale stress (SGS) models
for large-eddy simulation (LES) that are consistent with the numerical
scheme of the flow solver.  The method accounts for configurations
with two inhomogeneous directions and is applied to turbulent boundary
layers (TBL) under adverse pressure gradients (APG). To overcome the
limitations of linear eddy-viscosity closures in complex flows, we
adopt a non-Boussinesq SGS formulation along with a
dissipation-matching training loss. A second improvement is the
integration of a multi-task learning strategy that explicitly promotes
monotonic convergence with grid refinement, a property that is often
absent in conventional SGS models. \textit{A posteriori} tests show
that the proposed model improves predictions of the mean velocity and
wall-shear stress relative to the Dynamic Smagorinsky model (DSM),
while also achieving monotonic convergence with grid refinement.
\end{abstract}

\section{Introduction}

Wall-modeled large-eddy simulation (WMLES) has emerged as a popular
tool for high-Reynolds-number flow simulations in industrial
applications, offering a favorable balance between accuracy and
computational efficiency compared to direct numerical simulation (DNS)
and wall-resolved large-eddy simulation (WRLES)
\citep{kawaiWallmodelingLargeEddy2012,
  boseWallModeledLargeEddySimulation2018,
  maniPerspectiveStateAerospace2023}. In this framework, SGS models
are employed to account for the effects of unresolved small-scale
eddies, while wall models are used to bypass the prohibitive cost of
resolving near-wall motions.

Conventional SGS models typically rely on physical approximations and
empirical parameter tuning \citep{jimenezWhereAreWe1998,
  sagautLargeEddySimulation2006}. These approaches are often limited
by their underlying assumptions, which can lead to significant
inaccuracies in complex flow scenarios
\citep{gocLargeEddySimulation2021,
  agrawalNonBoussinesqSubgridscaleModel2022,
  huWallmodelledLargeeddySimulation2023,
  hayatWallModeledLargeEddySimulation2024}. A related limitation is
the common practice of developing wall and SGS closures as separate
components, which can introduce inconsistencies and compounding errors
in WMLES~\citep{lingWallmodeledLargeeddySimulation2022,
  arranzBuildingblockflowComputationalModel2024}. In addition,
discretization errors are frequently neglected when coarse grids are
used in practical simulations, further degrading predictive
accuracy~\citep{bae2017towards, lozano-duranErrorScalingLargeeddy2019,
  gocCertificationAnalysisLargeEddy2023,
  lozano-duranMachineLearningBuildingblockflow2023}. Finally, a
persistent challenge in WMLES is non-monotonic grid convergence of the
solution~\citep{larssonLargeEddySimulation2016,lozano-duranErrorScalingLargeeddy2019}. This
issue often stems from the interplay between SGS modeling and
numerical errors~\citep{lozano-duranErrorScalingLargeeddy2019,
  lozano-duranPerformanceWallModeledBoundaryLayerConforming2022,
  hu2024grid}, which can cause predictions to deteriorate even as the
mesh is refined.

While machine-learning-based (ML-based) SGS models have been developed
to address the limitations of conventional
closures~\citep{duraisamy2021perspectives, sanderse2024scientific,
  cho2025perspective}, early attempts often underperformed in
\textit{a posteriori} tests, especially on the coarse grids typical of
WMLES. This is primarily because they focused on \textit{a priori}
training using filtered or coarsened high-fidelity DNS data
\citep{gamaharaSearchingTurbulenceModels2017,
  xieArtificialNeuralNetwork2019,
  maulikSubgridModellingTwodimensional2019,
  parkNeuralnetworkbasedLargeEddy2021,
  liuInvestigationNonlocalDatadriven2021, kang2023neural,
  kim2024large, yang2024artificial, li2025mixed}. However, it is well
established that success in \textit{a priori} testing is only weakly
correlated with \textit{a posteriori} performance because filtered DNS
data are generally inconsistent with LES, except under very specific
conditions~\citep{lund2003use, bae2017towards, bae2018dns,
  bae2022numerical, huang2025consistency}. For this reason, recent
efforts have shifted toward \textit{a posteriori} training strategies.

\textit{A posteriori} training integrates the SGS model directly into
the WMLES solver at the target grid resolution, explicitly accounting
for numerical errors and ensuring consistency between training and
deployment. However, adoption has been slow due to the substantial
algorithmic and implementation complexity of these methods. Proposed
approaches include adjoint-based
methods~\citep{macartEmbeddedTrainingNeuralnetwork2021,
  sirignanoDeepLearningClosure2023,
  yuanAdjointbasedVariationalOptimal2023}, ensemble-based data
assimilation~\citep{monsEnsemblevariationalAssimilationStatistical2021,
  wangEnsembleDataAssimilationbased2023}, differentiable
solvers~\citep{list2022learned, shankar2025differentiable,
  fan2026diff}, reinforcement
learning~\citep{novatiAutomatingTurbulenceModelling2021, kim2022deep,
  beckDiscretizationConsistentClosureSchemes2023, kurz2025harnessing},
probabilistic models~\citep{EPHRATI2025114234}, continuous data
assimilation~\citep{ephrati2025continuous}, and the $\tau$-orthogonal
method~\citep{hoekstraReducedDatadrivenTurbulence2024,
  hoekstra2026reduced}. Despite their promise, widespread application
remains limited by high computational cost and the difficulty of
integrating these techniques into production WMLES solvers.

To date, a compelling ML-based SGS model leveraging data assimilation
(at least in terms of demonstrated applicability) is the
building-block flow model (BFM) proposed
by~\cite{lingWallmodeledLargeeddySimulation2022} and
\cite{arranzBuildingblockflowComputationalModel2024}. This framework
jointly learns an SGS closure and a wall model. The SGS component is
formulated as an optimization problem that matches DNS statistics
under a prescribed numerical scheme. The model trained via this
approach has demonstrated good performance across a broad range of
configurations, from canonical turbulent channel flows and pipes to
more complex geometries such as the Gaussian bump, the NASA Juncture
Flow, and the Common Research Model. However, the original BFM
data-assimilation method is restricted to configurations with only one
inhomogeneous direction, and it was also reported that certain
datasets involving separated flows could not be assimilated.

To address these challenges, \citet{ling2025numerically} proposed a
cost-effective data-assimilation-based approach for SGS modeling for
developing improved versions of the BFM. The method generates training
data using statistically nudged WMLES, in which a nudging term drives
the simulation toward prescribed target statistics. A
multiple-instance learning strategy is then used to regress the SGS
model against the resulting nudging forcing. In this paper, we extend
this framework to TBLs subject to APG, a configuration with two
inhomogeneous directions.  To accommodate the richer physics, we
generalize the model from a scalar eddy-viscosity closure to a
non-Boussinesq tensorial SGS formulation, which has been shown to
perform well in complex flows~\citep{moin1993new,
  lozano-duranInformationtheoreticFormulationDynamical2022,
  agrawalNonBoussinesqSubgridscaleModel2022}. During training, we also
enforce constraints on model dissipation, a statistical requirement
for SGS closures \citep{meneveau1994statistics, moser2021statistical}
that has been shown to be essential for capturing smooth-body
separation~\citep{iyer2024efficient}. Finally, we demonstrate improved
grid-refinement behavior, with enhanced convergence properties of the
resulting SGS model.

The paper is organized as follows. Section~\ref{sec:method} describes
the APG TBL case used for training and testing, the nudging approach
for flows with two inhomogeneous directions, and the SGS model and
machine learning formulation.  Section~\ref{sec:results} presents the
\textit{a posteriori} results and evaluates the effectiveness of the
convergence-enhancement strategy. Finally, Section~\ref{sec:dis}
discusses future improvements, and Section~\ref{sec:conc} concludes
the paper.

\section{Methods} \label{sec:method}



\begin{figure}
    \centering
    \includegraphics[width=\linewidth]{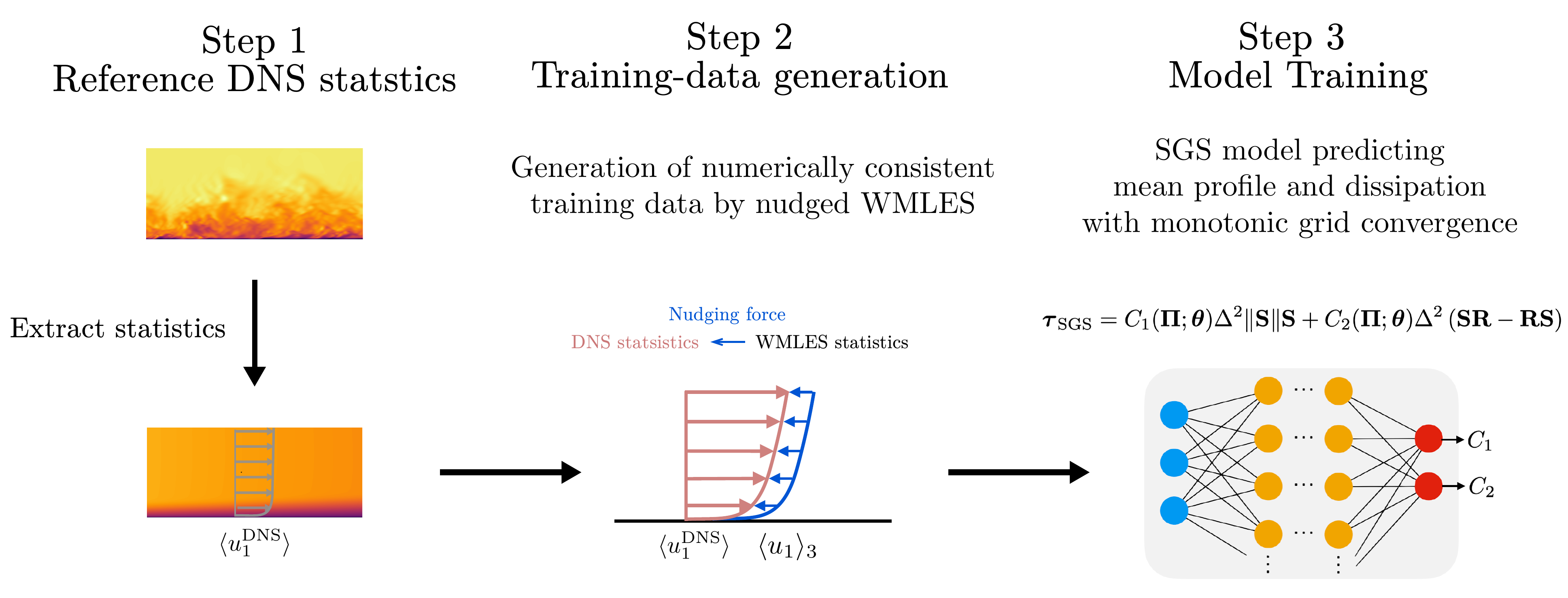}
    \caption{The three main steps for constructing the numerically
      consistent BFM SGS model.}
    \label{fig:schematic}
\end{figure}

The procedure for constructing the SGS model consists of three main
steps as shown in Figure~\ref{fig:schematic}:
\begin{enumerate}
\item \emph{Reference DNS statistics:} We extract the mean velocity
  profile from a DNS of a TBL under APG, which serves as the training
  target.
\item \emph{Training-data generation via nudging:} We run
  statistically nudged WMLES in the deployment solver to generate a
  numerically consistent target forcing that the SGS model must
  reproduce.
\item \emph{Model training:} We train the SGS model via supervised
  learning by minimizing a combined loss that matches the mean target
  forcing and the SGS dissipation obtained from the nudged simulations
  in step~2.
\end{enumerate}
This approach enables computationally efficient supervised training of
the SGS model while retaining key benefits typically associated with
(costly) reinforcement learning, such as consistency with the solver's
numerical schemes. In the following, we describe the process in more
detail. This study serves as a proof of concept, as we assimilate the
mean velocity profile from a single APG TBL case. Our longer-term goal
is to apply the approach to a larger set of DNS cases, on the order of
$\mathcal{O}(10)$--$\mathcal{O}(100)$.

\subsection{DNS of an APG TBL}

\begin{figure}[t]
    \centering
    \includegraphics[width=.8\linewidth]{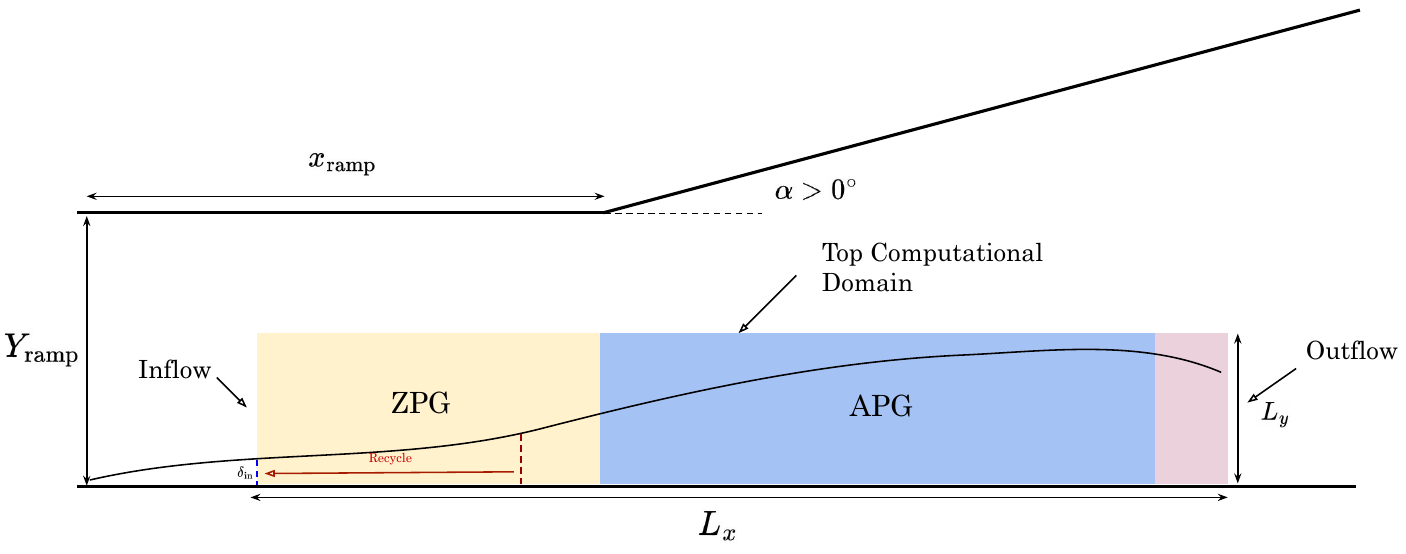}
    \caption{Schematic of the DNS setup of the APG TBL.}
    \label{fig:TBL_FIG}
\end{figure}

We train and test our model on a TBL case with moderate APG. The
schematic of the simulation is shown in Figure~\ref{fig:TBL_FIG}. The
adverse pressure gradient is introduced by prescribing a ceiling with
a virtual straight ramp that deflects the freestream upward. The case
considered in this study features a ramp angle $\alpha=5^{\circ}$, and
inflow momentum thickness Reynolds number
$Re_{\theta}=U_{\infty}\theta/\nu=670$.

The computational domain spans $L_x = 1200\theta_{in}$, $L_y =
200\theta_{in}$, and $L_z = 200\theta_{in}$ in the streamwise,
wall-normal, and spanwise directions, respectively, where
$\theta_{in}$ represents the momentum thickness at the inlet. The
inflow conditions are generated by prescribing a mean velocity profile
from a zero-pressure-gradient TBL at a matching Reynolds number
\citep{sillero2013one}. Velocity fluctuations, extracted from a
downstream recycling station, are superimposed on the mean profile
located at $x_{\text{recycle}} = 100 \theta_{in}$ using quasi-periodic
conditions.

At the domain exit, a convective outlet boundary condition is
implemented as $\partial \boldsymbol{u}/\partial t + U_{\infty}
\partial \boldsymbol{u}/\partial x = \boldsymbol{0}$
\citep{Pauley1990}, modified by small corrections to ensure global
mass conservation \citep{Simens2009}. The spanwise direction is
treated as periodic, while the top boundary utilizes an inviscid
potential-flow solution. This approach assumes a virtual wall located
above the domain, modeled via a pure potential source. The wall-normal
velocity at the upper boundary is defined as:
\begin{equation}
    v\big|_{L_y} = \frac{U_{\infty} L_y \left[ (x_{\text{ramp}} -
        x_{\text{source}})^{2} + L_y^{2}
        \right]}{\left(x_{\text{ramp}} - x_{\text{source}}\right)
      \left[ (x - x_{\text{source}})^{2} + L_y^{2} \right]},
\end{equation}
where the source location $x_{\text{source}}$ is derived by requiring
the potential flow to be parallel to the linear ramp at
$x_{\text{ramp}}$. This location represents the intersection of the
ramp line with the horizontal wall:
\begin{equation}
    x_{\text{source}} = x_{\text{ramp}} - Y_{\text{ramp}}\tan(\alpha).
\end{equation}
To maintain irrotational flow at the top boundary, the streamwise
velocity is determined by enforcing zero spanwise vorticity:
\begin{equation}
    \omega_z \big|_{L_y} = \frac{\partial v\big|_{L_y}}{\partial x} - \frac{\partial u\big|_{L_y}}{\partial y} = 0.
\end{equation}
Additionally, the spanwise velocity shear $\partial w/\partial y$ is
set to zero at this boundary.

The incompressible Navier-Stokes equations are solved using
DNS. Spatial derivatives are approximated using a second-order central
finite difference scheme on a staggered grid
\citep{orlandi2000fluid}. For temporal integration, a third-order
Runge-Kutta scheme \citep{wray1990minimal} is employed in conjunction
with a fractional-step method \citep{kim1985application}. The solver
has been validated in prior investigations of turbulent channel flows
\citep{bae2018turbulence, lozano2020non}, zero-pressure-gradient
turbulent boundary layers \citep{towne2023database}, and transitional
boundary layers \citep{Lozano2018_PSE}.

\begin{figure}[t]
    \centering
    \begin{subfigure}[b]{0.45\textwidth} 
        \includegraphics[width=\linewidth]{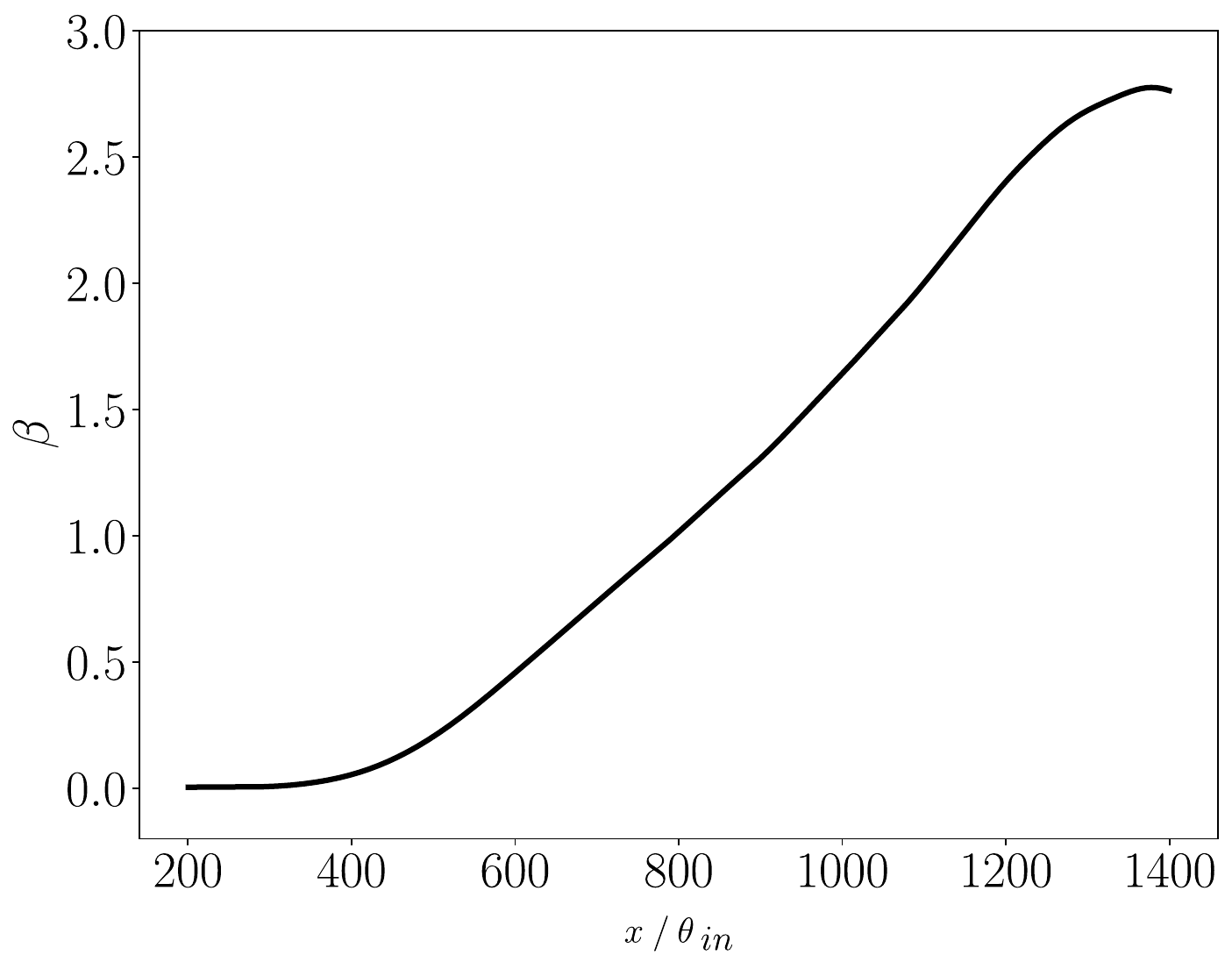}
        \caption{}
    \end{subfigure}
    \hfill
    \begin{subfigure}[b]{0.45\textwidth} 
        \includegraphics[width=\linewidth]{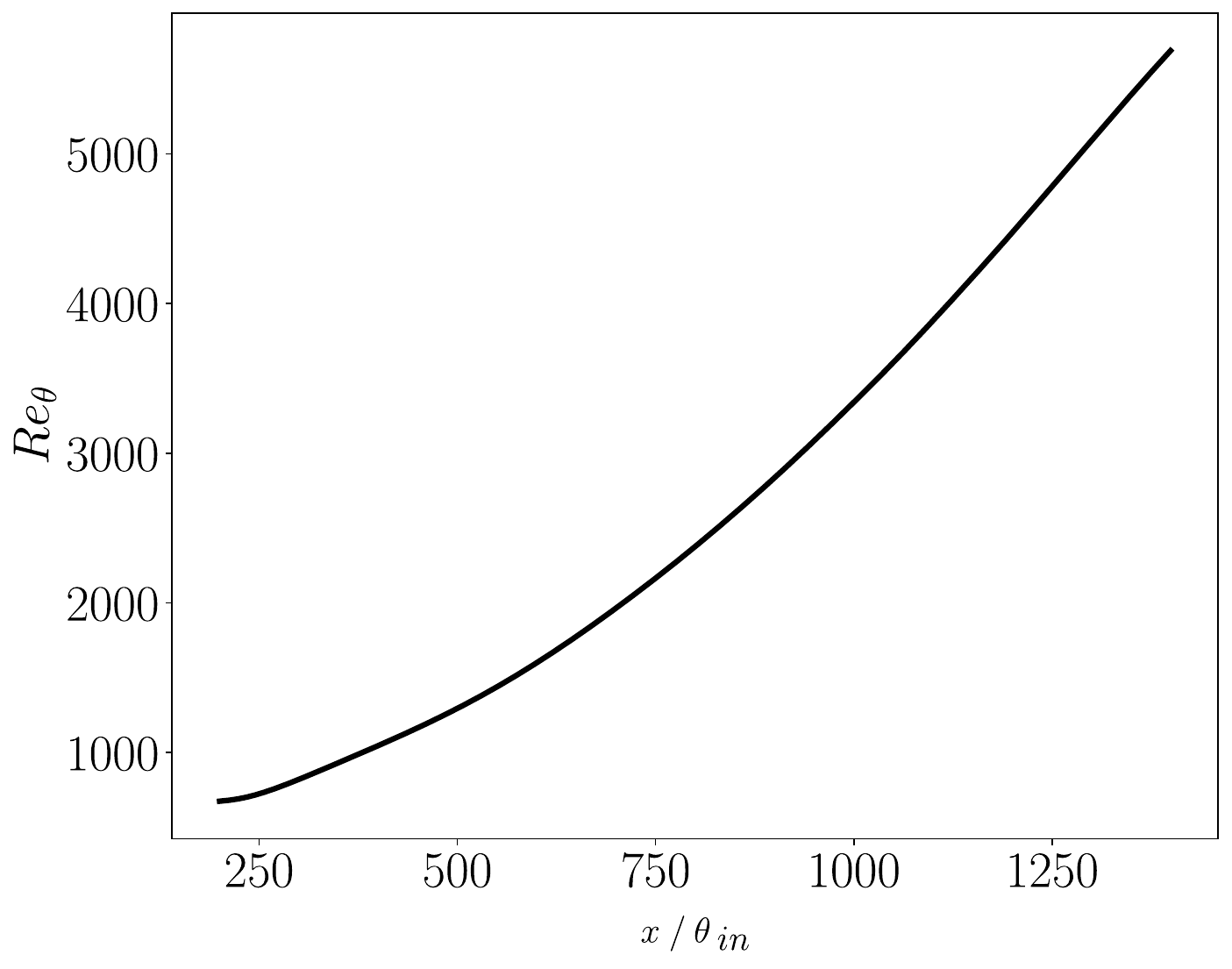}
        \caption{}
    \end{subfigure}
    \caption{Some properties of the APG TBL computed by DNS are used
      for training and testing. (a) Clauser parameter $\beta$. (b)
      Momentum-thickness Reynolds number $Re_{\theta}$.  }
    \label{fig:TBL5}
\end{figure}

Some properties of the APG TBL are shown in Figure~\ref{fig:TBL5}. The
Clauser pressure-gradient parameter, $\beta =
(\delta^*/\tau_w)\,\mathrm{d}P/\mathrm{d}x$, ranges from 0 to 2.8,
where $\delta^*$ is the displacement thickness, $\tau_w$ is the
wall-shear stress, and $\mathrm{d}P/\mathrm{d}x$ is the streamwise
freestream pressure gradient. The momentum-thickness Reynolds number
increases from 500 at the inlet to 5,500 near the outlet.

\subsection{Training data generation via nudging}

\begin{figure}[t]
    \centering
    \includegraphics[width=0.7\linewidth]{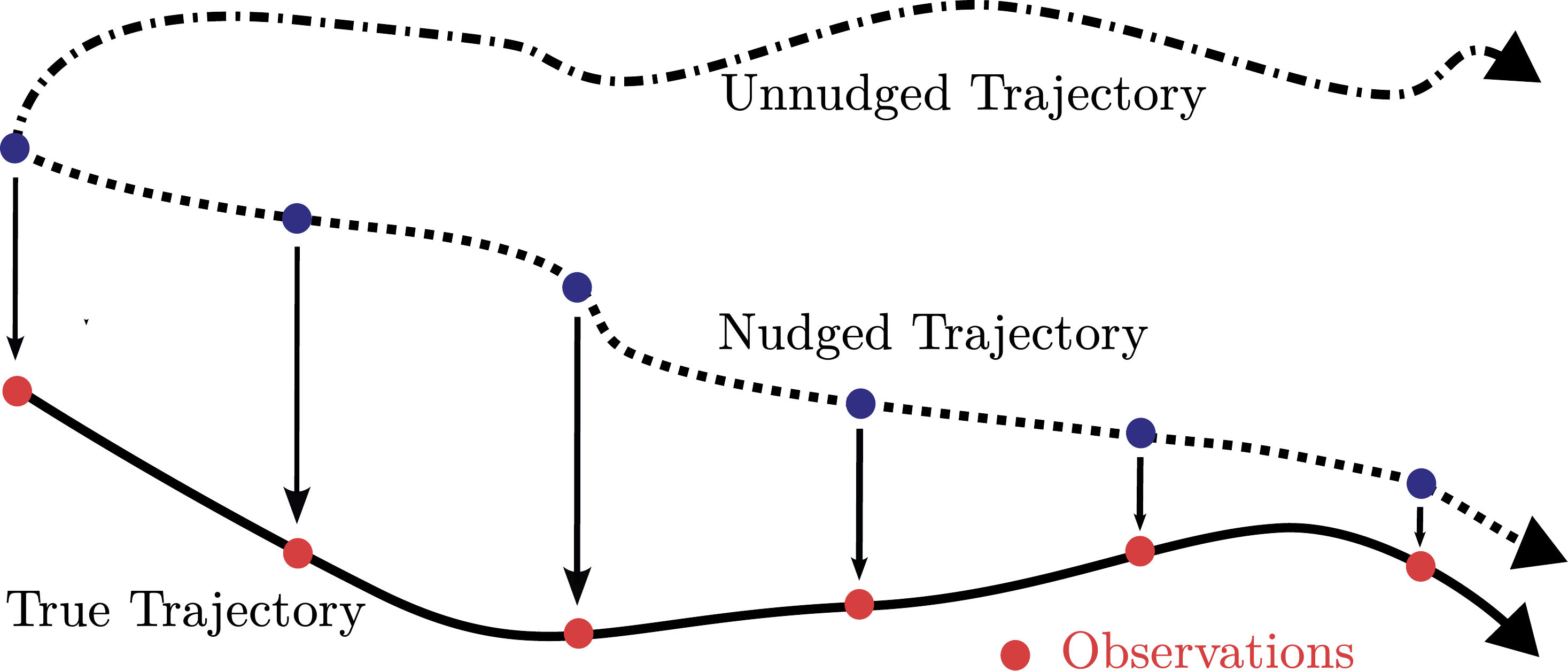}
    \caption{Schematic of nudging. Compared to the unnudged
      trajectory, nudging pushes the system to better match the true
      trajectory. The red circles are observations that are assumed to
      be the truth. Adapted from
      \citet{aschDataAssimilationMethods2016}.}
    \label{fig:nudging}
\end{figure}

The fundamental principle of nudging for a dynamical system is
illustrated in Figure~\ref{fig:nudging}. Mathematically, this entails
adding a relaxation term to the governing equations that nudges the
instantaneous flow state toward an observed reference state, which is
treated as the truth. Note, however, that this approach may not be
well suited for practical WMLES for several reasons. First, the notion
of a ``true trajectory'' is not well defined, since different
coarse-graining procedures applied to DNS would generally yield
different reference trajectories. Moreover, WMLES is chaotic, and
forcing the instantaneous trajectory to closely track a reference
trajectory is typically not possible without highly intrusive forcing.
Finally, even if WMLES is successfully nudged to a reference
trajectory, no SGS model can, in general, reproduce the required
forcing without error, owing to the fundamental information loss
inherent to coarse-grained
systems~\citep{lozano-duranInformationtheoreticFormulationDynamical2022}.
Because information from the small scales is irretrievably lost, there
is, in general, no solution with negligible error.

In the present work, rather than assimilating the full coarse-grained
trajectory, we assimilate only selected flow statistics. This offers
several advantages: the target is well defined; the statistics are
typically the quantities of practical interest; an SGS model that
matches these statistical constraints can, in general, be discovered;
and no costly space--time DNS data are required beyond the statistics,
which may also be obtained from experiments.
The statistical variant of nudging is achieved by adding a corrective
term to the right hand side of the LES momentum equations:
\begin{equation}
  \frac{\delta \mathbf{u}}{\delta t} + \mathbf{u} \cdot \nabla
  \mathbf{u} = -\frac{1}{\rho}\nabla p + \nabla \cdot ( 2\nu
  \mathbf{S} ) + \nabla \cdot \boldsymbol{\tau}_{\text{SGS}}^{\text{base}} +
  \mathbf{F}_{\text{nudging}},
  \label{eq:ns_nudging}
\end{equation}
where $\mathbf{u}$ is the resolved (LES) velocity field, 
$\delta /\delta t$ denotes the discrete time marching of the solver, $\nabla$ is the discrete nabla operator, 
$\nu$ is the
kinematic viscosity, $\boldsymbol{\tau}_{\text{SGS}}^{\text{base}}$ is
the baseline SGS stress tensor, $\mathbf{S}$ is the resolved
rate-of-strain tensor and $\mathbf{F}_{\text{nudging}}$ is the
correction term representing nudging, which is adjusted based on the
statistics to be assimilated. We define
\begin{equation}
\mathbf{F}_{\text{target}} = \nabla \cdot
\boldsymbol{\tau}_{\text{SGS}}^{\text{base}} +
\mathbf{F}_{\text{nudging}}
\end{equation}
as the forcing term our SGS model aims to approach to predict the
target statistics.  The DSM is used as the baseline SGS closure during
the training data generation, although other options are also
possible.  The selection of the DSM is motivated by previous reports
of its robust performance in complex cases \citep{goc2021large,
  goc2024wind}.

In our previous work~\citep{ling2025numerically}, we showed how the
nudging works for a channel-like simulation with only one
inhomogeneous direction. To apply the approach to a TBL simulation
with two inhomogeneous directions, we have
\begin{equation}\label{eq:nudging_final}
  \mathbf{F}_{\text{nudging}} = - \alpha_n (\langle u_1\rangle_{3} -
  \langle u_1^{\mathrm{DNS}} \rangle)\delta_{i1},
\end{equation}
where $\mathbf{u} = [u_1, u_2, u_3]$, with indices 1, 2, and 3
denoting the streamwise, wall-normal, and spanwise directions,
respectively, in a Cartesian coordinate system; $\delta_{ij}$ is the
Kronecker delta; the operator $\langle \cdot \rangle_{3}$ denotes an
average over the spanwise direction; $\alpha_n$ is a scalar that sets
the strength of the nudging forcing set to the minimum value such that
the relative error in the nudged mean velocity profile is below 1\%;
and $\langle u_1^{\mathrm{DNS}}\rangle$ is the DNS streamwise mean
velocity (averaged over the homogeneous directions and time).

We generate the training data for $\mathbf{F}_{\text{target}}$ using
the nudging formulation above implemented in the same WMLES solver
that is later used for \textit{a posteriori} simulations.  This
guarantees consistency with the numerical discretization of the flow
solver.  In our prior work~\citep{ling2025numerically}, we employed a
finite-difference code. Here, we implement the full training and
testing workflow within the GPU-accelerated solver \texttt{charLES}
(Cadence, Inc.). Our motivation for selecting this solver is to enable
the application of the SGS model to cases with arbitrary geometric
complexity. The \texttt{charLES} code is a compressible,
second-order-accurate finite-volume solver that employs
low-dissipation spatial discretization schemes. The solver leverages a
Voronoi-diagram-based meshing strategy to generate high-quality
unstructured grids and has been extensively validated across a broad
range of flow configurations~\citep{bres2017unstructured,
  goc2021large, goc2024wind, goc2025studies}.

Another important aspect of training data generation involves using a
pre-trained wall model throughout the nudging phase. This ensures that
the resulting flow fields incorporate the specific boundary-condition
effects associated with that wall model, allowing the nudging force,
and hence the SGS model, to learn wall-model-induced
behaviour~\citep{lingWallmodeledLargeeddySimulation2022,
  arranzBuildingblockflowComputationalModel2024}. Here, we use the
ML-based wall model BFM-WM-v2 developed by \cite{ling2025general} for
both data generation and \textit{a posteriori} testing. We select this
wall model because it exhibits low internal error (i.e., the error
induced when a wall model is provided with the correct outer-flow
inputs). 

\begin{figure}[t]
    \centering
    \includegraphics[width=.9\linewidth]{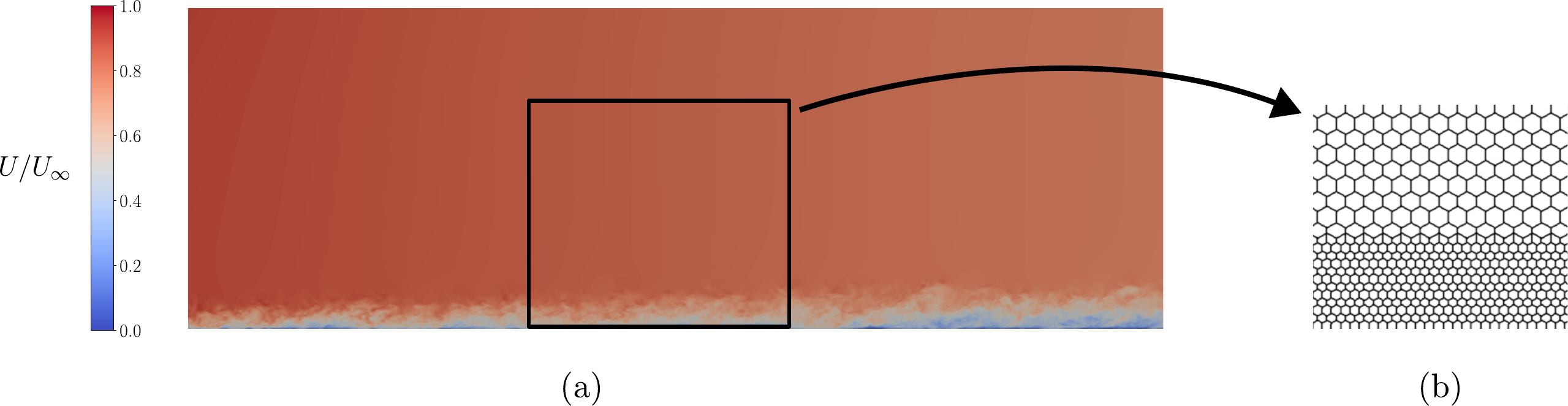}
    \caption{(a) The contour plot of the normalized instantaneous
      streamwise velocity at the center plane of the computational
      domain for the data generation of APG TBL case with a coarse
      grid. (b) The Voronoi grid in the black rectangular region shown
      in the left plot.}
    \label{fig:TBL_schematic}
\end{figure}

Training data are generated for the nudged APG TBL at three grid
resolutions, as summarized in Table~\ref{tab:grids}. This setup allows
the model to learn how to adjust its output across different grid
resolutions and to enforce monotonic grid convergence. An
instantaneous plot of the normalized streamwise velocity on the
coarsest grid is shown in Figure~\ref{fig:TBL_schematic}.
\begin{table}[t]
    \centering
    \begin{tabular}{ccccc}
       \hline
       Points per BL thickness        & Coarse & Medium & Fine  \\
       \hline
       Mean value    & 14 & 20 & 25  \\
       \hline
       Range interval    & [8, 24] & [10, 29] & [12, 36] \\
       \hline
    \end{tabular}
    \caption{Statistics for the number of grid points across the
      boundary layer thickness for three grid resolutions.}
    \label{tab:grids}
\end{table}

In summary, the nudged WMLES is performed using the \texttt{charLES}
solver for the APG TBL configuration. This setup uses the wall model
BFM-WM-v2, which is characterized by low internal error, together with
the DSM baseline SGS closure, to produce numerically consistent
datasets for training purposes. Finally, training data is generated
for three grid resolutions.

\subsection{SGS model formulation}

We adopt a non-Boussinesq tensorial formulation for the SGS model,
consistent with recent studies reporting improved
performance~\citep{lozano-duranInformationtheoreticFormulationDynamical2022,
  agrawalNonBoussinesqSubgridscaleModel2022, zhou2024sensitivity,
  zhou2025effect}. The original expansion for the SGS stress tensor
proposed by \citet{lundParameterizationSubgridscaleStress1992}
contains five distinct terms. Here, we seek a balance between
simplicity and accuracy by retaining a second term that accounts for
the rotation tensor, $\mathbf{R}$. The resulting model reads
\begin{equation}
\boldsymbol{\tau}_{\text{SGS}}=C_1(\mathbf{\Pi} ; \boldsymbol{\theta})
\Delta^2 \lVert\mathbf{S}\rVert \mathbf{S}+C_2(\mathbf{\Pi} ;
\boldsymbol{\theta})
\Delta^2\left(\mathbf{S}\mathbf{R}-\mathbf{R}\mathbf{S}\right),
\label{eq:sgs}
\end{equation}
where $\Delta$ is the characteristic grid size of a given control
volume, and $C_1$ and $C_2$ are parameterized by a dual-output
artificial neural network (ANN) with parameters
$\boldsymbol{\theta}$. The ANN takes as inputs the variables
$\mathbf{\Pi}$, which are dimensionless to ensure that the model is
invariant under changes of units. For the ANNs, we employ a
multi-layer perceptron architecture consisting of four hidden layers,
each with twenty neurons. The rectified linear unit is utilized as the
activation function for all hidden layers. The dimensionless inputs
$\mathbf{\Pi}$ are constructed from the dimensional variables:
\begin{equation}
  \mathbf{Q} = \{I_1, I_2, I_3, I_4, I_5, \nu, \Delta \},
  \label{eq:inv}
\end{equation}
where $I_i$, $i=1,\dots,5$, are local invariants of $\mathbf{S}$ and
$\mathbf{R}$~\citep{lundParameterizationSubgridscaleStress1992}:
\begin{equation}
  \begin{aligned}
    I_1 &= \operatorname{tr}(\mathbf{S}^2), & I_2 &= \operatorname{tr}(\mathbf{R}^2), \\
    I_3 &= \operatorname{tr}(\mathbf{S}^3), & I_4 &= \operatorname{tr}(\mathbf{S}\mathbf{R}^2), \\
    I_5 &= \operatorname{tr}(\mathbf{S}^2 \mathbf{R}^2).
  \end{aligned}
\end{equation}
The use of these invariants to construct $\mathbf{\Pi}$ ensures that
the model is equivariant under translations and rotations, and
invariant under Galilean transformations.

To determine the optimal dimensionless inputs, we use the
information-theoretic Buckingham-$\pi$ theorem from
\citet{yuan2025dimensionless}. This approach leverages
information-theoretic principles to identify the input variables that
maximize the predictive power of the output regardless of the model structure (from analytical functions to neural networks). The core premise is that
the performance of a model is fundamentally bounded by the information
shared between inputs and outputs~\citep{lozano2022information},
defined formally by Shannon information~\citep{shannon1948}.

Since explicit target values for $C_1$ and $C_2$ are unavailable, in
contrast to standard supervised learning, we cannot directly measure
the mutual information of $\mathbf{\Pi}$ with respect to these
coefficients. Instead, we optimize a lower-bound proxy to select the
optimal inputs. Since predicting the target forcing remains the primary
goal, this selection is based on the relationship between the
dimensionless mean forcing and the mean inputs. Consider the
functional dependency chain:
\begin{equation}
    C_1, C_2 \rightarrow \mathbf{F}_{\text{target}} \rightarrow
    \langle \mathbf{F}_{\text{target}}\rangle,
\end{equation}
where each arrow represents a deterministic mathematical
operation. The data-processing inequality~\citep{cover1999elements}
states that such operations cannot increase mutual
information. Consequently, we have the following relationships:
\begin{equation}
    I(\langle \mathbf{\Pi}\rangle;\langle \mathbf{F}_{\text{target}}\rangle)\leq
    I(\mathbf{\Pi};\langle \mathbf{F}_{\text{target}}\rangle) \leq
    I(\mathbf{\Pi}; \mathbf{F}_{\text{target}} ) \leq I(\mathbf{\Pi}; C_1,C_2),
\end{equation}
where $I(\cdot;\cdot)$ denotes the mutual information. While the
information-theoretic Buckingham-$\pi$ theorem implies that
$I(\mathbf{\Pi}; C_1,C_2)$ is the target quantity for optimization,
data availability restricts us to computing the leftmost term,
$I(\langle \mathbf{\Pi}\rangle;\langle \mathbf{F}_{\text{target}}
\rangle)$. Therefore, by maximizing this term, we effectively optimize
a lower bound on $I(\mathbf{\Pi}; C_1,C_2)$. Applying this procedure
yields the following set of optimal dimensionless inputs:
\begin{equation}
    \mathbf{\Pi} = \left[ 
    \frac{I_1^{1/2}\Delta^2}{\nu}, \quad
    \frac{I_3}{I_1^{3/2}}, \quad
    \frac{I_5}{I_1^{2}}, \quad
    \frac{I_4}{I_1^{3/2}}, \quad
    \frac{I_3 I_4}{I_1^{3}} 
    \right].
\end{equation}

\subsection{Loss function and training details}\label{subsec:nn}

The loss function has two components: one to promote accurate
prediction of $\left\langle
\mathbf{F}_{\text{target},i}\right\rangle_3$ and another to match the
dissipation of the baseline and learned SGS models. The problem is
formulated using a multiple-instance learning approach analogous to
\cite{ling2025numerically}. Samples at each grid resolution are
partitioned into $K=3$ datasets, such that the $i$-th dataset contains
$N_i$ samples. The first component of the loss for the $i$-th dataset,
which targets the nudging forcing, is defined as
\begin{align}
    \mathcal{L}_{\text{forcing},i} &= \frac{1}{N_i}
    \left\lVert \left\langle \mathbf{F}_{\text{pred}}\right\rangle_{3} - \left\langle
    \mathbf{F}_{\text{target}}\right\rangle_3 \right\rVert^2,\\
    \mathbf{F}_{\text{pred}} &= \nabla \cdot \boldsymbol{\tau}_{\text{SGS}},
    %
\end{align}
where $\mathbf{F}_{\text{pred},i}$ is the forcing predicted by the
learned model, and the subscript $i$ denotes samples constrained to
belong to the $i$-th dataset.  Note that the target forcing
$\mathbf{F}_{\text{target}}$ does not, in general, conserve
momentum, since it may act as a source or sink. In contrast, the
learned model $\boldsymbol{\tau}_{\text{SGS}}$ is momentum-conserving,
because it enters the dynamics only through its divergence.

The second component of the loss function ensures that the model
reproduces the dissipation induced by the SGS stresses. We first note
that the second tensor-expansion term in Eq.~\ref{eq:sgs} does not
contribute to mean SGS dissipation or production, due to symmetry. A
close inspection of the energy equation in Appendix~\ref{app:en_eq}
indicates that the nudging term should induce additional
dissipation. However, order-of-magnitude arguments show that the
dissipation associated with the nudging term is negligible compared
with that of the eddy-viscosity term (Appendix~\ref{app:order}). We
therefore neglect the nudging contribution and define the dissipation
loss for the $i$-th dataset as
\begin{align}
    \mathcal{L}_{\text{dissipation},i} &= \frac{1}{N_i}
    \left\lVert \left\langle \varepsilon_{\text{pred}}\right\rangle_3 - \left\langle
    \varepsilon_{\text{target}}\right\rangle_3 \right\rVert^2,\\
    \varepsilon_{\text{pred}} &= \boldsymbol{\tau}_{\text{SGS}} : \mathbf{S}, \\
    \varepsilon_{\text{target}} &= \boldsymbol{\tau}_{\text{SGS}}^{\text{base}} : \mathbf{S},
\end{align}
where $:$ denotes the inner product. The final loss function is
\begin{equation}\label{eq:totalloss}
\mathcal{L}_{\text{total}} = \sum_{i=1}^{K=3} w_i(\beta_{d,i}
\mathcal{L}_{\text{dissipation},i} + \beta_{f,i}
\mathcal{L}_{\text{forcing},i}),
\end{equation}
where weights $w_i$ provides a mechanism to adjust the contribution of
each dataset based on its perceived importance and $\beta_{d,i}$ and
$\beta_{f,i}$ are weights adjusted dynamically to ensure that the loss components are dimensionally consistent and comparable in magnitude. Specifically, these are
defined as the inverse of the running mean of their respective loss
components, $\beta_{k,i} = 1/\mu_{k,i}$ for $k \in \{d, f\}$ and $i\in \{1,2,3 \}$. At each
training epoch $m$, the running mean $\mu_{k,i}$ is updated via an
exponential moving average as follows:
\begin{equation}\label{eq:running_mean}
\mu_{k,i}^{(m)} = (1 - \gamma) \mu_{k,i}^{(m-1)} + \gamma \mathcal{L}_{k,i}^{(m)},
\end{equation}
where $\gamma \in (0, 1]$ is a smoothing momentum hyperparameter, set
  to be 0.9 in this work.

Regarding the weights $w_i$, we consider two cases: (i) equal weights
across datasets, and (ii) resolution-dependent weights. The latter
aims to assess whether emphasizing finer-grid datasets improves
convergence under grid refinement, rather than assigning the same
importance to lower-resolution cases. The weights used in each case
are summarized in Table~\ref{tab:weights}.
\begin{table}
    \centering
    \begin{tabular}{ccccc}
       \hline
       & $w_{\text{Coarse}}$ & $w_{\text{Medium}}$ & $w_{\text{Fine}}$  \\
       \hline
       Unequal Weights & 1 & 10 & 100  \\
       \hline
       Equal Weights    & 1 & 1 & 1 \\
       \hline
    \end{tabular}
    \caption{Two versions of the model with different weight configurations 
    assigned to datasets at each resolution.}
    \label{tab:weights}
\end{table}

For each dataset, 30,000 individual streamwise stations are included,
with the individual station serving as the fundamental unit for data
splitting and sampling. The total dataset is partitioned into
training, validation, and testing subsets using a 7:1.5:1.5 ratio. The
model is implemented using PyTorch~\citep{paszke2019pytorch} and
optimized via the Adam algorithm~\citep{kingma2014adam} with an
initial learning rate of 0.0005 and decay rates of $0.9$ and
$0.999$. To ensure robust convergence, the learning rate is reduced by
a factor of 0.975 if the validation loss fails to decrease for 5
consecutive epochs. Furthermore, an early stopping criterion
terminates the training process if no improvement is observed for 20
epochs.

\section{Results} \label{sec:results}

\subsection{A posteriori test at the finest grid size} \label{subsec:apost}

We present \textit{a posteriori} results for the highest grid
resolution used during training (the fine grid in
Table~\ref{tab:grids}), using the DSM with the same wall model
BFM-WM-v2 as the benchmark. Results obtained with other grid
resolutions were found to be qualitatively similar.

\begin{figure}[t]
    \centering
    \begin{subfigure}[b]{0.64\textwidth} 
        \includegraphics[width=\linewidth]{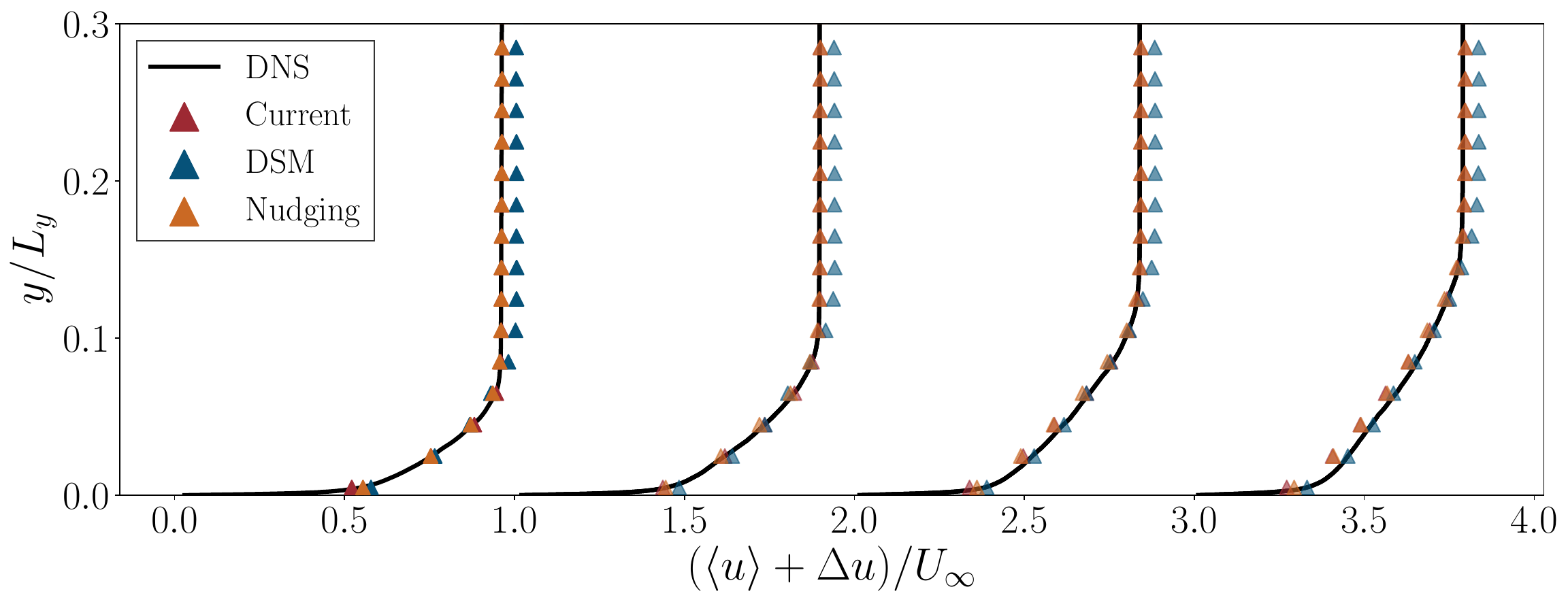}
        \caption{}
        \label{fig:sub:u}
    \end{subfigure}
    \hfill
    \begin{subfigure}[b]{0.29\textwidth} 
        \includegraphics[width=\linewidth]{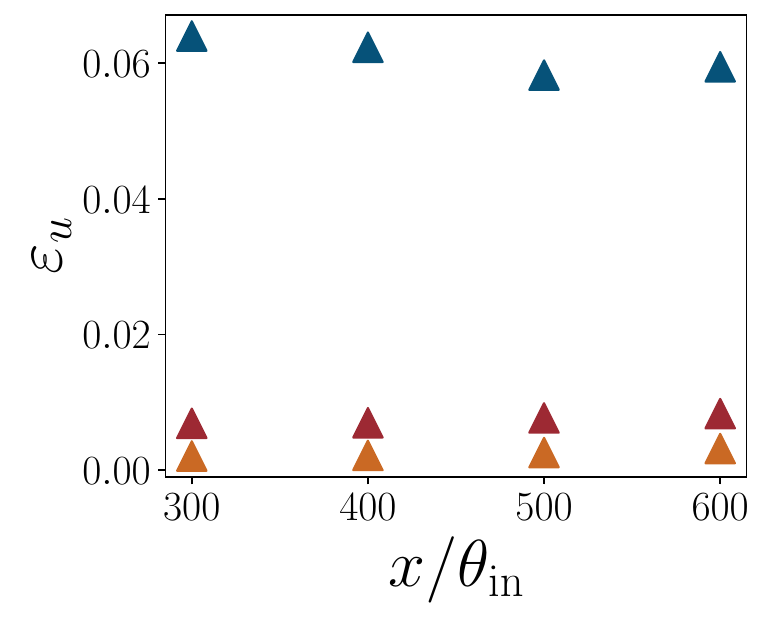}
        \caption{}
        \label{fig:sub:u_error}
    \end{subfigure}
    \caption{\textit{A posteriori} testing of the current SGS model
      and the DSM both with BFM-WM-v2 in WMLES of an APG TBL. The
      configuration corresponds to a turbulent boundary layer
      developing under a top ramp inclined at $5^{\circ}$.  (a) Mean
      velocity profiles at four streamwise locations, $x/L_y = 3, 4,
      5,$ and $6$ ($x/\theta_{\text{in}} = 300, 400, 500, 600$) as a function of the wall normal direction $y$. 
      Profiles are vertically shifted by $\Delta u /
      U_{\infty} = x/L_y - 3$ for clarity. (b) Integrated relative error
      $\varepsilon_u$ at these stations.}\label{fig:u}
\end{figure}

The predicted mean streamwise velocity profiles (here denoted by $u$ for simplicity) at selected streamwise
locations are presented in Figure~\ref{fig:sub:u} . To quantify the
predictive accuracy across the domain, we compute the integrated
error, $\varepsilon_u$, defined as
\begin{equation}
\varepsilon_u = \frac{1}{\delta}\int_0^{\delta} \frac{| \langle u \rangle -
  \langle u^{\text{DNS}} \rangle|}{U_e} \,\mathrm{d}y,
\end{equation}
where $\delta$ denotes the boundary-layer thickness and $U_e$ is the
edge velocity. As shown in Figure~\ref{fig:sub:u_error}, the current
SGS model significantly outperforms the standard DSM and yields
results comparable to the optimal nudged prediction.

\begin{figure}[t]
\centering
    \begin{subfigure}[b]{0.5\textwidth}
        \includegraphics[width=.932\linewidth]{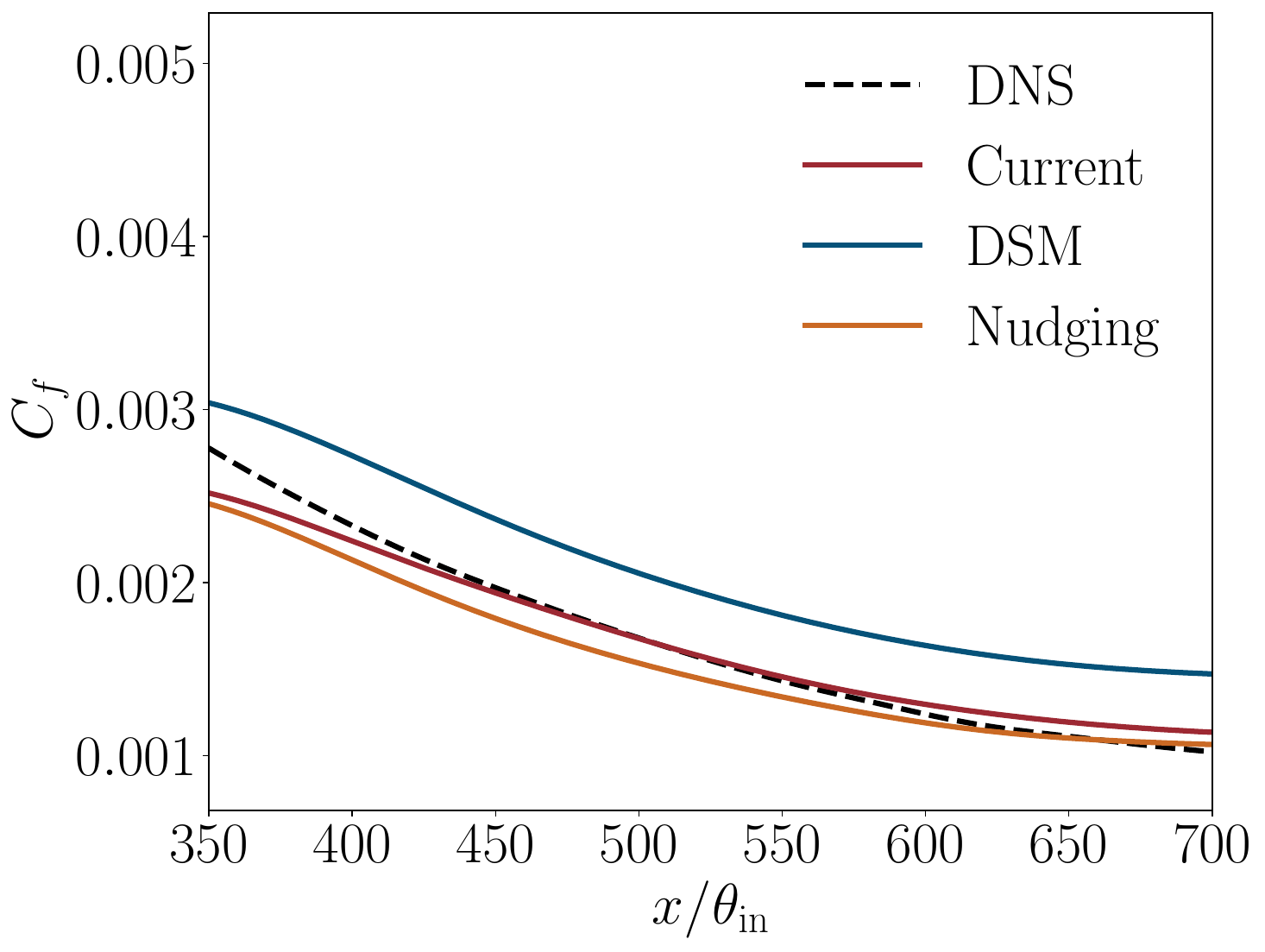}
        \caption{}
        \label{fig:sub:tauw_deg5_comp}
    \end{subfigure}
    \hfill
    \begin{subfigure}[b]{0.453\textwidth}
        \includegraphics[width=\linewidth]{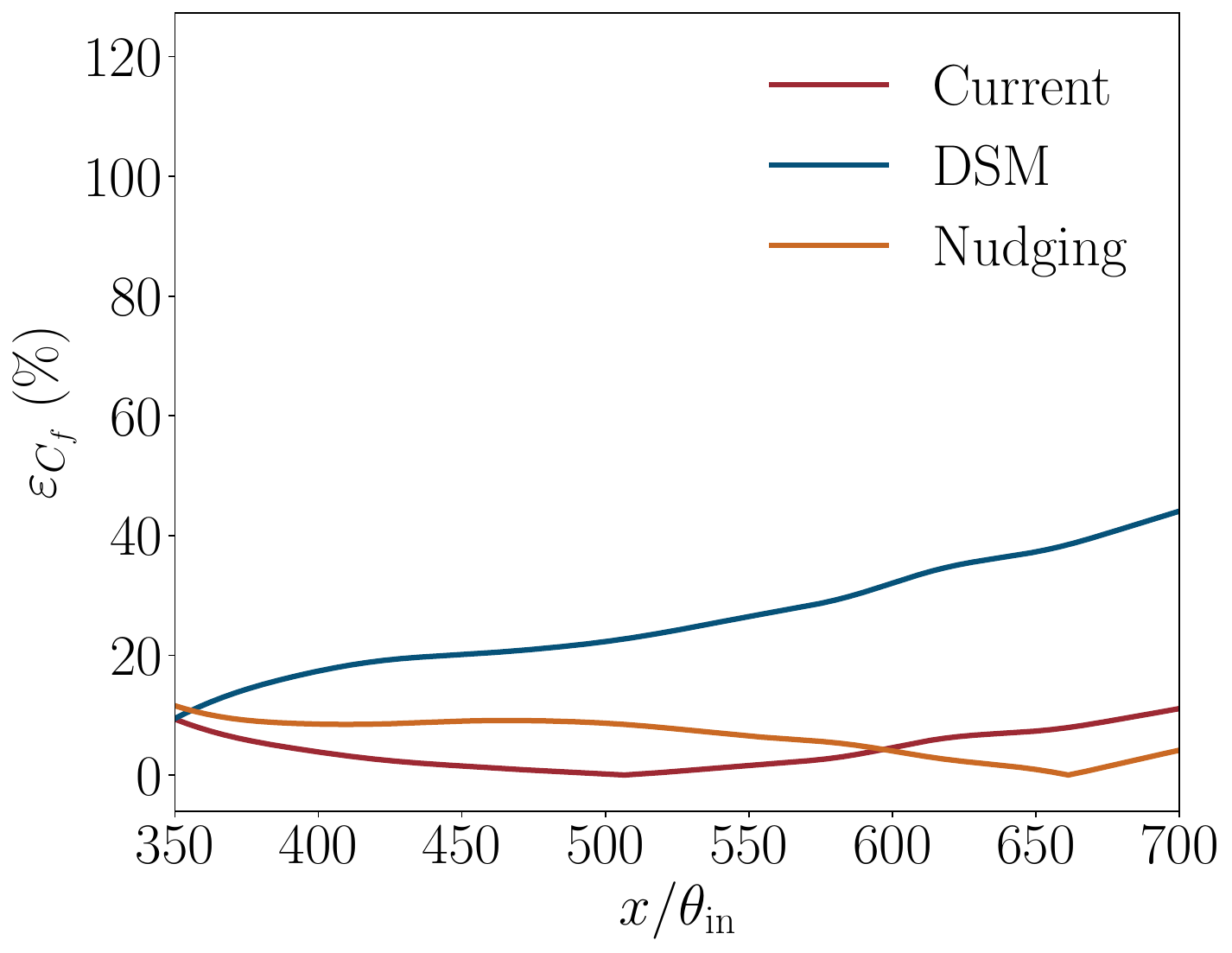}
        \caption{}
        \label{fig:sub:tauw_deg5_comp_err}
    \end{subfigure}
    \caption{(a) Streamwise distribution of the predicted
      skin-friction coefficient $C_f$ by the current SGS model and the
      DSM both with BFM-WM-v2 in WMLES of an APG TBL.  (b)
      Relative error $\varepsilon_{C_f} = (C_{f,\text{pred}} -
      C_{f,\text{DNS}}) / C_{f,\text{DNS}}$.}
    \label{fig:tauw_deg5_apot}
\end{figure}

Figure~\ref{fig:tauw_deg5_apot} illustrates the prediction of the
skin-friction coefficient. As anticipated, the discrepancies in the
mean velocity profile produced by the DSM translate into substantial
errors in the wall-shear stress. Conversely, the current SGS model
closely tracks the reference data. In certain regions, the current
model appears to outperform even the nudged simulation, i.e., the
theoretical ideal forcing. This counterintuitive result is likely
attributable to partial error cancellation between the SGS model and
the wall model, creating an illusion of superior performance compared
to the nudged flow.

\subsection{Convergence behavior with grid refinement} \label{subsec:conv}

\begin{figure}[t]
    \centering
    \includegraphics[width=0.6\linewidth]{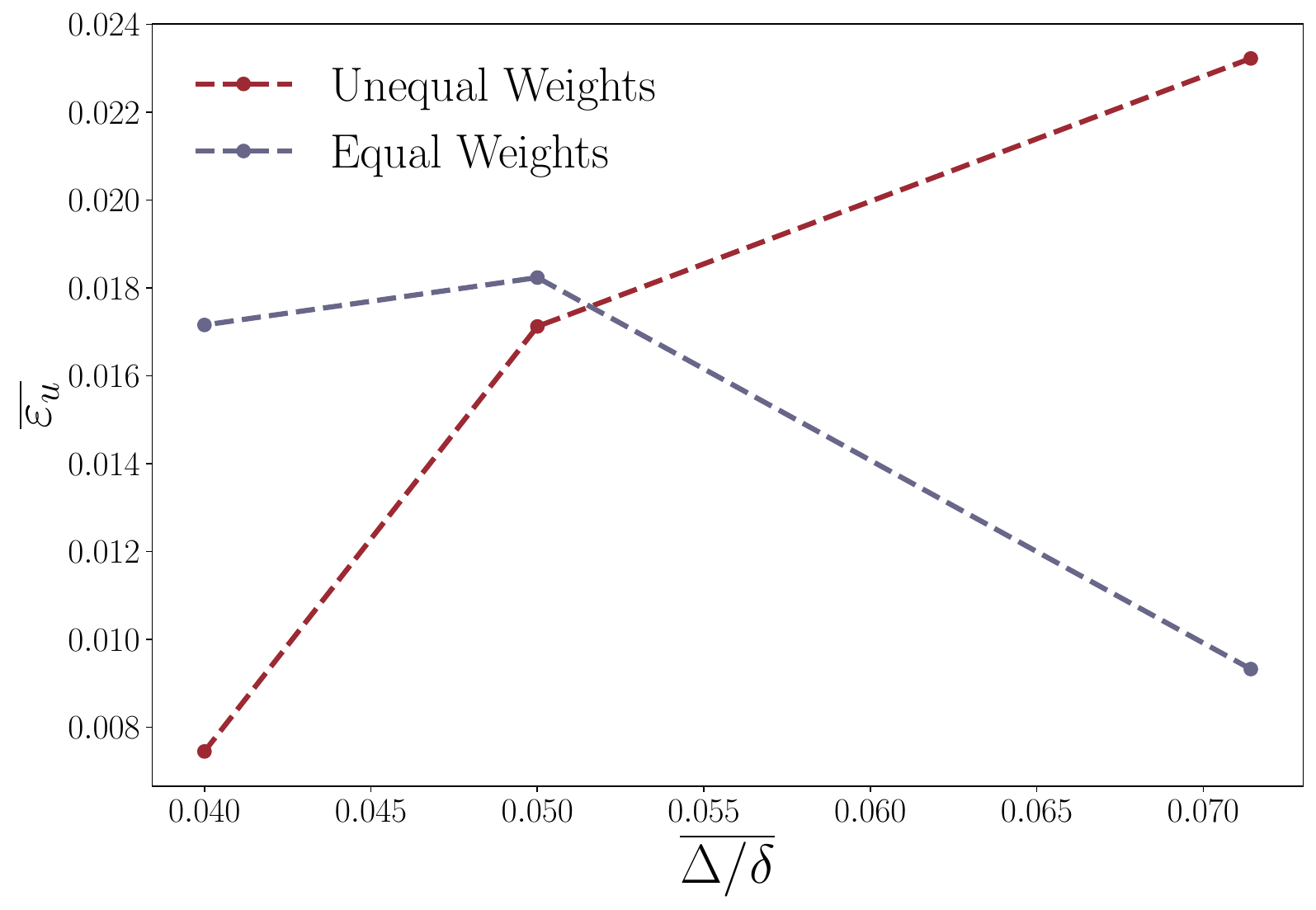}
    \caption{Comparison of mean velocity error at different grid
      resolutions using SGS models trained with equal weights and
      unequal weights.}
    \label{fig:conv}
\end{figure}

The convergence behavior of the model is analyzed in
Figure~\ref{fig:conv}. We adopt the integrated error in the mean
velocity, $\overline\varepsilon_u$, as the primary metric for
assessing convergence. The reported error is averaged over the region
of interest $x/\theta_{\text{in}} \in [300,700]$.

Results are shown for the three grid resolutions used in training. We
first consider the trained SGS model with equal weights across the
three grid sizes. For this unweighted model, the coarsest grid
paradoxically yields the smallest error, while the error increases at
finer resolutions. This is a trend contrary to the desired monotonic
grid-convergence behavior. By contrast, employing the proposed
weighting strategy recovers a consistent convergence trend. Although
the finest resolution remains under-resolved and not yet fully
converged, the multi-task strategy of assigning distinct weights to
different grid resolutions clearly promotes the monotonic grid
convergence desired in WMLES.

\section{Discussion} \label{sec:dis}

\subsection{General nudging scheme}

The current nudging-based data assimilation approach constructs a
one-dimensional forcing term for the streamwise velocity. While
effective for canonical boundary layers and channels at moderate mean
pressure gradients, this formulation restricts the applicability of
the method to flows whose mean statistical state varies strongly in
more than one direction, such as separated flows. Other examples
include square-duct configurations or wing--body junctions, where
secondary mean motions are
critical~\citep{pirozzoli2018turbulence}. In such cases, the nudging
force becomes vector-valued, correcting not only the streamwise
momentum deficit but also driving the secondary cross-stream vortices
(the $v$ and $w$ components) that are often under-resolved in standard
WMLES. It remains to be shown whether this straightforward extension
is robust for these more complex configurations.

Furthermore, the scope of assimilation can be expanded beyond
first-order moments (mean velocity) to include higher-order turbulence
statistics, such as resolved Reynolds stresses or the turbulent
kinetic energy dissipation rate. Assimilating dissipation is
particularly attractive for training SGS models, as it directly
constrains the energy transfer from resolved to subgrid
scales. However, this introduces a coupled optimization challenge:
forcing terms designed to correct second-order statistics must be
introduced without contaminating the mean-flow forcing. It is not yet
clear whether this extension can be implemented without degrading the
original performance.

\subsection{SGS model tensor form}

The current model formulation incorporates two terms from the complete
five-term integrity basis derived by
\citet{lundParameterizationSubgridscaleStress1992}. This parsimonious
choice was made to enhance specific structural properties without
directly altering the SGS dissipation characteristics. However, for
flow regimes exhibiting more complex stress anisotropies---such as
corner flows, junctions, or systems with strong rotation---the
remaining terms in the expansion may become necessary. Arbitrarily
including all five terms, however, increases both computational cost
and the risk of overfitting the training data. To address this, future
work will explore principled feature-selection techniques to identify
an optimal subset of tensor terms.

\subsection{Scalability via automated multi-task learning}

The current work establishes a proof of concept for using
resolution-dependent loss weighting to enforce grid convergence in
data-driven SGS models. However, this demonstration was limited to a
single flow configuration with a manually tuned weighting schedule. To
ensure the method is robust enough for general-purpose deployment,
future work must assess its performance across a broader training
database, such as the diverse collection of APG TBL cases with varying
Reynolds numbers and pressure-gradient histories presented by
\citet{ling2025general}.

Scaling this approach to multiple flow cases exposes a critical
limitation of the current strategy: reliance on fixed, manually
assigned weights. As the number of training cases and grid levels
increases, the hyperparameter space for these weights becomes
prohibitively large to tune by hand. Consequently, integrating
automated multi-task learning algorithms becomes essential.

Promising approaches exist, such as the Pareto-optimal training
strategy for turbulence modeling demonstrated by
\citet{liu2025towards}, or the homoscedastic uncertainty weighting
proposed by \citet{kendall2018multi}, which dynamically balances loss
terms based on intrinsic task uncertainty. However, applying these
methods directly to WMLES presents a unique challenge. Standard
algorithms typically aim to balance performance across all tasks or
focus on ``easy'' tasks first. In contrast, our physical constraint is
hierarchical: error reduction on finer grids must be prioritized over
coarser grids to satisfy consistency requirements. Determining how to
modify existing automated objectives to strictly enforce this specific
convergence hierarchy, rather than merely balancing errors, remains an
open and nontrivial question for future research.

\section{Conclusions} \label{sec:conc}

This study presents a data-driven, numerically consistent SGS model
for WMLES. The model is constructed using a statistical formulation of
nudging, which is applied to a turbulent boundary layer under an
adverse pressure gradient. The results demonstrate the robustness of
this data-assimilation-based training strategy for flows with
increased complexity.

In addition to statistical nudging, we introduced two primary
modifications to our previous data-driven SGS modeling approach to
improve performance and numerical convergence. First, we adopted a
non-Boussinesq tensorial formulation and added a dissipation
constraint to the loss function. The resulting model outperforms the
standard dynamic Smagorinsky model coupled with the same wall model
for the case examined. Second, we developed a multi-task learning
method to control the convergence behavior of the data-driven SGS
model. By weighting the training datasets according to grid
resolution, we enforce consistent error reduction with grid
refinement. This strategy helps address a persistent challenge in
WMLES, where interaction errors among wall models, numerics, and SGS
models often produce non-monotonic convergence.

\section*{Acknowledgement}
This work was supported by the National Science Foundation (NSF) under
grant number \#2317254 and NSF CAREER \#2140775.  The project is also
supported by an Early Career Faculty grant from NASA’s Space
Technology Research Grants Program (grant \#80NSSC23K1498).  The
authors acknowledge the MIT Office of Research Computing and Data for
providing high performance computing resources that have contributed
to the research results reported within this paper. This work also
used the DeltaAI and Delta system at the National Center for
Supercomputing Applications: Services and Support
(ACCESS) program, which is supported by National Science Foundation
grants \#2138259, \#2138286, \#2138307, \#2137603, and \#2138296.

\appendix

\section{SGS energy equation for streamwise nudged WMLES governing equations}\label{app:en_eq}
Suppose that we are only nudging in the streamwise direction and we
set it in the x direction. Then the SGS energy equation is

\begin{multline}
    \frac{\partial k_{sgs}}{\partial t} + \bar{u}_j \frac{\partial k_{sgs}}{\partial x_j} = 
    -\underbrace{\tau_{ij} \bar{S}_{ij}}_{\text{SGS production}\ \varepsilon_{sgs}} -\underbrace{\alpha \overline{(u_1 - \bar{u}_1)F_1}}_{\text{Nudging production}\ S_{k_{sgs}}}\\
    + \nu \frac{\partial^2 k_{sgs}}{\partial x_j \partial x_j} 
    - \frac{\partial}{\partial x_j} \left( \frac{1}{2}\overline{u'_i u'_i u'_j} + \frac{1}{\rho}\overline{p' u'_j} \right) 
    - \nu \overline{\frac{\partial u'_i}{\partial x_j} \frac{\partial u'_i}{\partial x_j}},
\end{multline}
where the first two terms correspond to the production of SGS energy production from the SGS term ($\varepsilon_{sgs}$) and nudging term ($S_{k_{sgs}}$) and therefore the dissipation from the resolved energy. $F_1 =\alpha (\langle u_1\rangle_3 - \langle u_1^{\mathrm{DNS}} \rangle)$ is the magnitude of $F_{\text{nudging}}$ in Eq.~\ref{eq:nudging_final}.
\section{Order-of-Magnitude analysis of the resolved energy dissipation: SGS vs. Nudging}\label{app:order}

We compare the magnitudes of $|\varepsilon_{sgs}|$ and $|S_{k_{sgs}}|$ in a typical WMLES with high Reynolds number. First, if $\varepsilon_{sgs}$ is only a function of $k_{sgs}$ and $\Delta$, we assume $\varepsilon_{sgs} = C_\varepsilon k_{sgs}^{3/2} / \Delta$ and the scaling $k_{sgs} \sim (u'_{sgs})^2$ holds. Then we have
\begin{equation}
|\varepsilon_{sgs}| \sim \frac{(u'_{sgs})^3}{\Delta}.
\end{equation}
Then for the Nudging term, we assume that the relevant velocity scale is outer scale $U$ and outer length scale $L$. Mark the SGS fluctuation $u_1-\bar{u}_1=\tilde{u}_1$. Then we have the magnitude $|S_{k_{sgs}}| \sim |\tilde{u}_1| \cdot |F_1| \cdot \rho_{\tilde{u}_1F_1}$, where $\rho_{\tilde{u}_1F_1}$ is the correlation coefficient. Since the SGS fluctuation scales with the deviation from the mean, $|\tilde{u}_1|\sim u'_{sgs}$, and $F_1\sim U^2/L$ by dimensional analysis, we have
\begin{equation}
|S_{k_{sgs}}| \sim u'_{sgs} \cdot \frac{U^2}{L} \cdot \rho_{\tilde{u}_1F_1}.
\end{equation}

Then we calculate the ratio between two terms:
\begin{equation}
\frac{|\varepsilon_{sgs}|}{|S_{k_{sgs}}|} \sim \frac{(u'_{sgs})^3 / \Delta}{u'_{sgs} \cdot (U^2/L) \cdot {\rho_{\tilde{u}_1F_1}}} = \left(\frac{u'_{sgs}}{U}\right)^2 \cdot \frac{L}{\Delta} \cdot \frac{1}{\rho_{\tilde{u}_1F_1}}.
\end{equation}
By assuming that the flow is in inertial subrange and therefore equating the dissipation from the outer large scale and inner scale, we have $(u'_{sgs}/U) \sim (\Delta/L)^{1/3}$. Using this scaling leads to 
\begin{equation}
\frac{|\varepsilon_{sgs}|}{|S_{k_{sgs}}|} \sim \left( \left(\frac{\Delta}{L}\right)^{1/3} \right)^2 \cdot \frac{L}{\Delta} \cdot \frac{1}{\rho_{\tilde{u}_1F_1}} = \left(\frac{\Delta}{L}\right)^{2/3} \cdot \frac{L}{\Delta} \cdot \frac{1}{\rho_{\tilde{u}_1F_1}}=\left(\frac{L}{\Delta}\right)^{1/3} \cdot \frac{1}{\rho_{\tilde{u}_1F_1}}.
\end{equation}
By scale separation assumption, we have $\Delta \ll L$ and the correlation between $\tilde{u}_1$ (decided by the small scale) and $F_1$ (decided by the large scale) is small. Therefore, it is reasonable to assume $\rho_{\tilde{u}_1F_1}\ll 1$. Putting all these together, we conclude that for a typical WMLES at high Reynolds number, we have
$$
|\varepsilon_{sgs}| \gg |S_{k_{sgs}}|.
$$

\bibliography{Reference}

\end{document}